\documentclass[5p,times,twocolumn]{elsarticle}

\usepackage{lineno,hyperref}
\modulolinenumbers[5]
\usepackage{graphicx}
\usepackage{booktabs}
\usepackage{siunitx}
\usepackage{amsmath}

\journal{Acta Astronautica}

\begin{document}

\begin{frontmatter}

\title{Suborbital measurements of atmospheric structure and cosmic radiation over the Mexican Plateau}

\author[uaemex]{U. Ochoa-Torrentera}
\author[aem]{R.\,A. V\'azquez-Robledo}
\author[uaemex]{J. Sumaya-Mart\'inez\corref{cor1}}

\cortext[cor1]{Corresponding author. Email: \href{mailto:jsm@uaemex.mx}{jsm@uaemex.mx}}

\address[uaemex]{Facultad de Ciencias, Universidad Aut\'onoma del Estado de M\'exico, Toluca, Estado de M\'exico, Mexico}
\address[aem]{Agencia Espacial Mexicana, Mexico}

\begin{abstract}
We report suborbital in situ measurements of atmospheric temperature, pressure, and ionizing cosmic radiation obtained during a stratospheric balloon mission over the Mexican Plateau. The balloon reached a maximum altitude of \SI{28.94}{km} above mean sea level, traversing the troposphere, tropopause, and lower stratosphere. Temperature and pressure profiles were recorded continuously during ascent and descent and were used to derive atmospheric density as a function of altitude. The measured thermal stratification enables a piecewise description compatible with the International Standard Atmosphere (ISA) framework for this region. Ionizing radiation measurements exhibit the expected altitude dependence of secondary cosmic radiation, including a maximum consistent with the Regener--Pfotzer maximum, with a peak dose rate of \SI{2.96}{\micro\gray\per\hour} at \SI{18.64}{km}. To the best of our knowledge, this is the only documented stratospheric balloon mission conducted in the State of Mexico that combines near-space atmospheric profiling with direct measurements of ionizing cosmic radiation up to altitudes approaching \SI{30}{km}.
\end{abstract}


\end{frontmatter}


\section{Introduction}

Stratospheric balloon platforms constitute a reliable and cost-effective approach for accessing the near-space environment, enabling in situ measurements in the altitude range between commercial aviation and low Earth orbit satellites. These platforms have been widely employed in atmospheric physics, aerospace engineering, and space radiation studies due to their ability to provide continuous vertical profiles with relatively low operational complexity and cost \cite{ISO2533,USSA1976,WallaceHobbs2006}.

The vertical thermodynamic structure of the Earth's atmosphere is commonly described by reference models such as the International Standard Atmosphere (ISA), which provides standardized profiles of temperature, pressure, and density as functions of altitude \cite{ISO2533,USSA1976}. These models are extensively used in aerospace applications, navigation, and preliminary mission design. However, ISA represents climatological mean conditions and does not explicitly account for regional variations associated with geographic location, surface elevation, or transient meteorological phenomena. Consequently, experimental measurements at regional scales are required to assess the applicability and limitations of standard atmospheric models for specific environments \cite{HoltonHakim2013}.

In addition to thermodynamic variables, the atmospheric radiation environment represents a critical aspect of near-space studies. Primary cosmic rays interacting with atmospheric nuclei generate cascades of secondary particles, producing an altitude-dependent ionizing radiation field. This radiation reaches a maximum intensity in the lower stratosphere, commonly referred to as the Regener--Pfotzer maximum \cite{RegenerPfotzer1935,Pfotzer1936,Carlson2014}. Characterization of this radiation profile is relevant for space weather research, radiation exposure assessment in aviation, and the design of suborbital and high-altitude instrumentation \cite{SheaSmart2000,BentonBenton2001}.

Stratospheric balloon experiments have played a central role in the experimental study of atmospheric cosmic radiation, complementing satellite-based and ground-based observations. Recent balloon campaigns have demonstrated the suitability of such platforms for radiation dosimetry and particle flux measurements in the stratosphere \cite{Mertens2016Dose,Mertens2016Overview,Gronoff2016}. Despite these advances, experimental datasets combining atmospheric profiling and in situ cosmic radiation measurements remain limited for many geographic regions.

The Mexican Plateau offers favorable conditions for suborbital atmospheric studies due to its high elevation and mid-latitude location. Launch sites in this region reduce the effective atmospheric column above the payload, facilitating access to the lower stratosphere. Nevertheless, to the best of our knowledge, no previously published stratospheric balloon mission conducted within the State of Mexico has reported combined measurements of temperature, pressure, and ionizing cosmic radiation up to altitudes approaching \SI{30}{km}.

In this work, we present the results of a stratospheric balloon mission launched from Aculco, Estado de M\'exico, designed to obtain vertical profiles of temperature, pressure, atmospheric density, and ionizing radiation from approximately \SI{2.5}{km} above mean sea level to a maximum altitude of \SI{28.94}{km}. The objectives of this study are: (i) to experimentally characterize the thermodynamic structure of the lower atmosphere over the Mexican Plateau; (ii) to compare the measured profiles with ISA predictions; and (iii) to analyze the altitude dependence of ionizing cosmic radiation in the troposphere and lower stratosphere. This mission represents, to the best of our knowledge, the only documented experiment of its kind in the State of Mexico and provides region-specific data relevant to atmospheric physics and near-space radiation studies \cite{ConacytUAEMex2018}.

\section{Experimental setup}

\subsection{Stratospheric balloon platform}

The experiment was conducted using a latex stratospheric balloon filled with helium. The balloon was launched from Aculco, Estado de M\'exico, at an initial altitude of approximately \SI{2.5}{km} above mean sea level. During ascent, the balloon expanded until reaching a burst altitude of \SI{28.94}{km}, after which the payload descended by parachute. Measurements were recorded continuously during both ascent and descent.

The payload was housed in a lightweight insulated structure designed to operate under low-pressure and low-temperature conditions. A global positioning system (GPS) provided altitude, geographic coordinates, and time stamps, enabling correlation of all measured parameters with altitude.

\subsection{Instrumentation}

Atmospheric temperature and pressure were measured using digital sensors integrated into the payload. Ionizing radiation was monitored using a radiation detection module sensitive to secondary cosmic-ray particles. Sensor data were acquired and logged by an onboard microcontroller-based data acquisition system. Power was supplied by batteries rated for low-temperature operation. Table~\ref{tab:instr} summarizes the instrumentation employed in the mission.

\begin{table}[t]
\centering
\caption{Instrumentation used in the stratospheric balloon mission.}
\label{tab:instr}
\begin{tabular}{@{}llll@{}}
\toprule
Parameter & Sensor type & Range & Purpose \\
\midrule
Temperature & Digital sensor & \SI{-40}{\celsius} to \SI{60}{\celsius} & Thermal profile \\
Pressure & Barometric sensor & \SIrange{300}{1100}{hPa} & Pressure--altitude relation \\
Ionizing radiation & Radiation detector & Relative dose / count rate & Cosmic radiation profile \\
Position \& altitude & GPS module & Global & Trajectory and altitude \\
Data acquisition & Microcontroller-based & --- & Data logging \\
\bottomrule
\end{tabular}
\end{table}

\section{Results}

\subsection{Temperature profile}

The measured temperature profile exhibits a stratified structure characteristic of the troposphere, tropopause, and lower stratosphere. In the troposphere, temperature decreases approximately linearly with altitude, with an average lapse rate of about \SI{-6.8}{\celsius\per\km}. A reduced gradient is observed in the tropopause region, followed by a positive gradient of approximately \SI{4.1}{\celsius\per\km} in the lower stratosphere.

\begin{figure}[t]
\centering
\includegraphics[width=\linewidth]{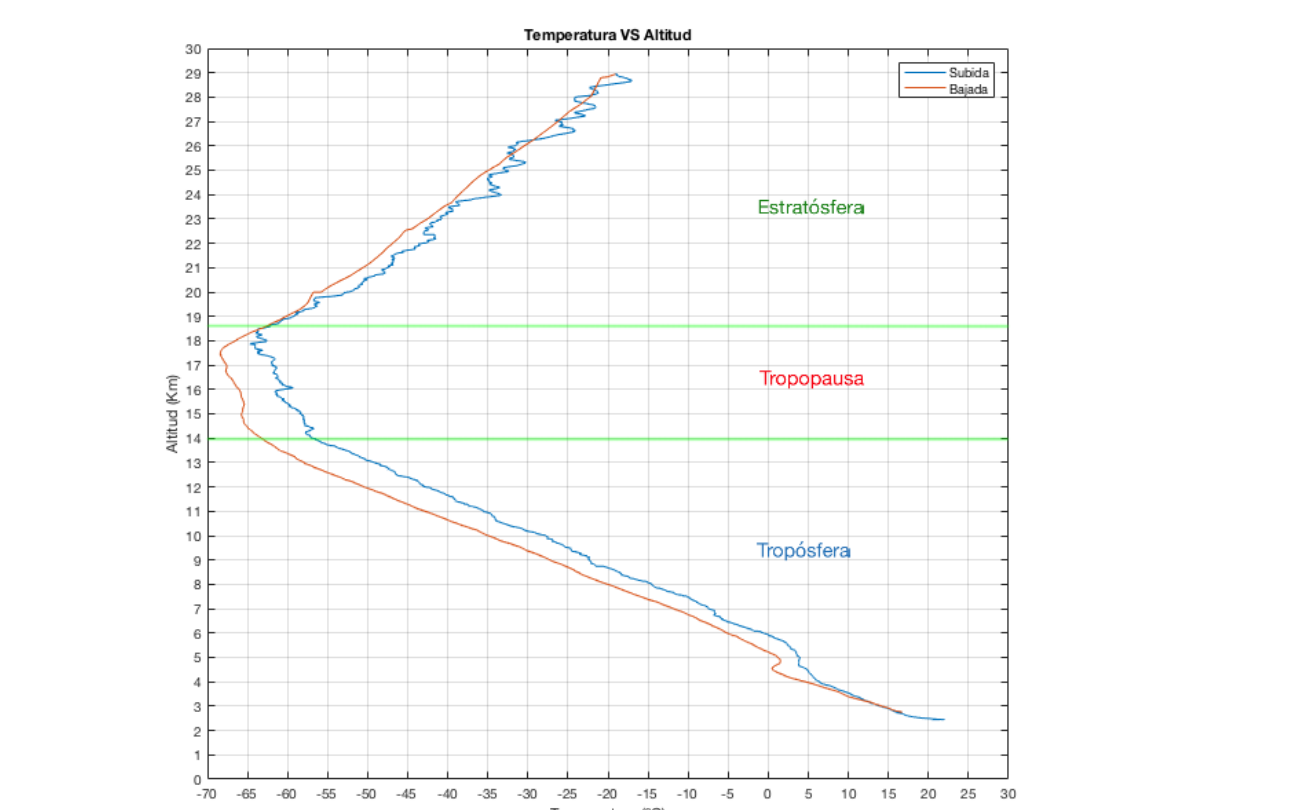}
\caption{Measured temperature profile during ascent and descent. The trajectory traverses the troposphere, tropopause, and lower stratosphere.}
\label{fig:temp}
\end{figure}

Temperature profiles obtained during ascent and descent show minor differences, which are attributed to variations in vertical velocity, horizontal displacement during flight, and temporal changes in atmospheric conditions. Figure~\ref{fig:temp} shows the measured temperature as a function of altitude during ascent and descent.

\subsection{Pressure and derived density}

Atmospheric pressure decreases monotonically with altitude throughout the flight. The measured pressure profile follows an exponential trend consistent with hydrostatic equilibrium. Using the measured temperature and pressure data and assuming ideal gas behavior, atmospheric density was derived as a function of altitude. The resulting density trend is consistent with ISA values across most of the altitude range studied, with modest deviations near the tropopause where atmospheric variability and temperature gradients are enhanced.

\subsection{Ionizing radiation measurements}

\begin{figure}[t]
\centering
\includegraphics[width=\linewidth]{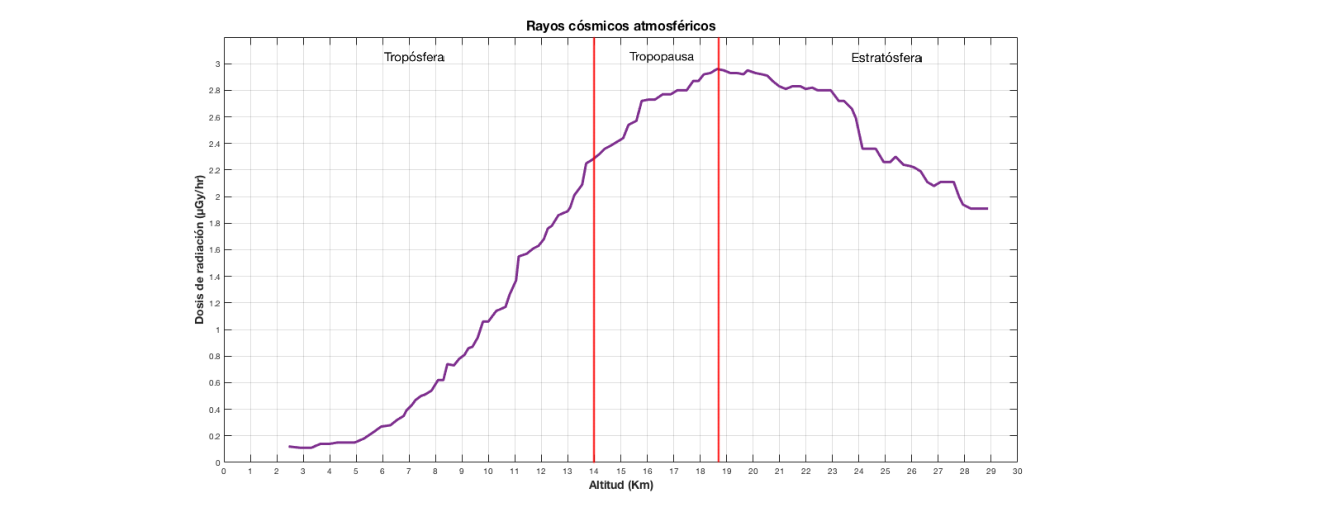}
\caption{Ionizing radiation dose rate versus altitude (ascent). The peak near \SI{18.6}{km} is consistent with the Regener--Pfotzer maximum.}
\label{fig:rad}
\end{figure}

Ionizing radiation measurements show a clear increase with altitude, reflecting the reduced atmospheric shielding and the development of secondary cosmic-ray cascades. The radiation signal increases steadily during ascent, reaching a maximum at an altitude of about \SI{18.64}{km}, consistent with the Regener--Pfotzer maximum, with a peak dose rate of \SI{2.96}{\micro\gray\per\hour}. Figure~\ref{fig:rad} shows the radiation dose rate as a function of altitude for the ascent profile.

\subsection{Trajectory overview}

\begin{figure}[t]
\centering
\includegraphics[width=\linewidth]{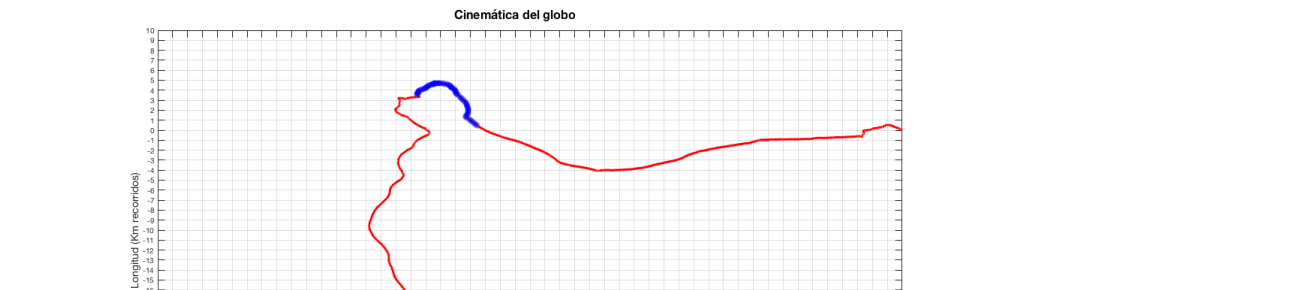}
\caption{Reconstructed balloon trajectory in a local reference frame (qualitative), showing horizontal drift during the flight.}
\label{fig:traj}
\end{figure}

The GPS record indicates horizontal displacement during ascent and descent, implying that the atmospheric profile combines measurements taken at different locations and times. Figure~\ref{fig:traj} shows the reconstructed trajectory in a local reference frame.

\section{Discussion}

The experimental results demonstrate that a stratospheric balloon platform can resolve the vertical thermodynamic structure of the lower atmosphere over the Mexican Plateau. The measured temperature gradients and pressure trends are broadly consistent with ISA predictions, supporting the applicability of standard atmospheric models at regional scales while highlighting the influence of local atmospheric conditions.

The radiation measurements provide an additional layer of scientific relevance, extending the mission beyond conventional meteorological soundings. The identification of an altitude-dependent increase and a maximum consistent with the Regener--Pfotzer maximum confirms the suitability of stratospheric balloons for near-space radiation studies and provides region-specific constraints for exposure assessment and model validation.

A significant contribution of this work is its regional character. To the best of our knowledge, this mission represents the only documented stratospheric balloon experiment conducted in the State of Mexico that combines atmospheric profiling with in situ measurements of ionizing cosmic radiation up to altitudes approaching \SI{30}{km} \cite{ConacytUAEMex2018}. This dataset therefore helps fill an observational gap for central Mexico.

The primary limitation of this study is the reliance on a single flight and a sensor suite oriented toward feasibility rather than metrology-grade calibration. Future campaigns incorporating repeated launches, improved detector characterization, and complementary radiation transport modeling would enable a more comprehensive characterization of the regional near-space environment \cite{BadhwarONeil2010}.

\section{Conclusions}

A stratospheric balloon mission conducted over the Mexican Plateau provided experimental measurements of temperature, pressure, atmospheric density, and ionizing radiation from about \SI{2.5}{km} to \SI{28.94}{km} above mean sea level. The measured thermodynamic profiles exhibit the expected atmospheric stratification and show overall agreement with the International Standard Atmosphere model within experimental uncertainties.

Ionizing radiation measurements reveal a clear altitude dependence and a maximum consistent with the Regener--Pfotzer maximum in the lower stratosphere. To the best of our knowledge, this mission constitutes the only reported experiment in the State of Mexico combining near-space atmospheric profiling with direct measurements of cosmic radiation up to altitudes approaching \SI{30}{km}. These results demonstrate the potential of stratospheric balloon platforms for atmospheric and radiation studies in near-space environments and provide region-specific data relevant to aerospace applications and space radiation research.

\section*{Acknowledgements}
The authors acknowledge the support of the Facultad de Ciencias (UAEM\'ex) and the Agencia Espacial Mexicana.

\bibliographystyle{elsarticle-num}
\bibliography{references}

\end{document}